\documentclass[preprint2]{aastex}

\slugcomment{}
\shorttitle{Detection of N$_{2}^{+}$ in Leonid Fireball}
\shortauthors{Abe et al.}

\begin{document}

\title{Detection of the N$_{2}^{+}$ First Negative System in a Bright Leonid Fireball}

\author{Shinsuke Abe}
\affil{Astronomical Institute of the Academy of Sciences,
298 Fricova, Ondrejov 25165, The Czech Republic;}
\email{avell@asu.cas.cz}

\author{Noboru Ebizuka}
\affil{Institute of Physical and Chemical Research,
Wako, Saitama 351-0198, Japan;}
\email{ebizuka@riken.jp}

\author{Hajime Yano}
\affil{Institute of Space and Astronautical Science, 
Japan Aerospace Exploration Agency, 
3-1-1 Yoshinodai, Sagamihara, Kanagawa 229-8510, Japan;}
\email{yano.hajime@jaxa.jp}

\author{Jun-ichi Watanabe}
\affil{National Astronomical Observatory of Japan, 
National Institutes of Natural Sciences, 
2-21-1 Osawa, Mitaka, Tokyo 181-8588, Japan;}
\email{jun.watanabe@nao.ac.jp}

\and

\author{Ji\v{r}\'{i} Borovi\v{c}ka}
\affil{Astronomical Institute of the Academy of Sciences,
298 Fricova, Ondrejov 25165, The Czech Republic;}
\email{borovic@asu.cas.cz\\ {\it Accepted on 2004 Nov. 30 for publication in} The
Astrophysical Journal Letters}

\begin{abstract}
An ultraviolet-visible spectrum between 300 and 450 nm of a cometary
meteoroid that originated from 55P/Tempel-Tuttle was investigated, and
its new molecules, induced by atmospheric interaction, were
discovered. The spectroscopy was carried out using an intensified
high-definition TV camera with a slitless reflection grating during the
2001 Leonid meteor shower over Japan. A best-fit calculation mixed with
atoms and molecules confirmed the first discovery of N$_{2}^{+}~
B^{2}\Sigma_{u}^{+} \rightarrow X^{2}\Sigma_{g}^{+}$ bands in the UV
meteor spectrum. The N$_{2}^{+}$ temperature was estimated to be 10,000
K with a low number density of 1.55 $\times$ 10$^{5}$ cm$^{-3}$. Such
unexpectedly strong ultraviolet emission, in particular for N$_{2}^{+}$
(1,0) at 353.4 nm, is supposed to be formed through the wide dimensions
of high-temperature regions caused by a large meteoroid. Spectroscopic
observations of reentry capsules will provide us with good opportunities
for confirming the discovered N$_{2}^{+}$.
\end{abstract}


\section{Introduction}

Spectroscopic observations of meteors reveal not only the chemical
composition of the cometary meteoroids but also the emission processes
of hypervelocity impacts in the atmosphere; these processes are
difficult to reproduce in laboratory experiments at present. Leonid
meteoroids that correspond to cometary grains from the comet
55P/Tempel-Tuttle have produced the best meteor shower because of its
high incident velocity at $\sim$72 km s$^{-1}$ among known annual meteor
showers and because of the bright flux of its meteors at $\sim$10,000
hr$^{-1}$.

Of particular interest is the question as to whether or not meteoroids
could have delivered organics and water to the early Earth
\citep{jenn00}. \citet{rietmeijer02} suggests that Leonid meteors are
large aggregates that might include precursors to the interplanetary
dust particles (IDPs) and that the survival of meteoritic compounds
through the atmospheric entry is feasible even if its at high-velocity
ablation. To determine whether large cometary grains contain mineral
water or trapped water in any forms, it is necessary to confirm the
presence of OH $A^{2}\Sigma^{+}\rightarrow X^{2}\Pi$ emission around a
wavelength of 310 nm. \citet{harvey77}, \citet{abe02,abe03} and
\citet{jenn02b} reported an excess of emission at 310 nm.

Here we report the discovery of N$_{2}^{+}~ B^{2}\Sigma_{u}^{+}
\rightarrow X^{2}\Sigma_{g}^{+}$ [N$_{2}^{+}$(1--)] in the wavelength
range of 320-450 nm meteor emission from a Leonid meteoroid through the
investigation of the OH $A$--$X$ (0,0) band. The N$_{2}^{+}$(1--) plasma
emission in meteors has been argued by \citet{millman71} and
\citet{mukhamednazarov77}. Meanwhile, \citet{jenn04a} found a
N$_{2}^{+}~ A^{2}\Pi_{u} \rightarrow X^{2}\Sigma_{g}^{+}$ Meinel band in
the range of 780-840 nm. Thus, our discovery is important for
identifying unknown meteor emissions in the ultraviolet region, in
particular, for understanding the variety of emission phases in meteors
and the delivery mechanisms of organic matter, content minerals, and
water.

\section{Observations and data reduction}

During the 2001 Leonid maximum, spectroscopic observations were carried
out using Image-Intensified High-Definition TV (II-HDTV) cameras in the
ultraviolet (UV), visible (VIS), and near-infrared (near-IR) wavelength
regions (250-700 nm). The II-HDTV system consisted of a UV image
intensifier ($\phi$ 18 -mm photocathode: S20), two relay lenses (f=50mm,
f/1.4), and an HDTV camera with a 2 mega pixels CCD. In order to focus
precise optical concentration on the wavelength range of 250-1000 nm, we
developed UV lenses of f=30 mm, f/1.2 with a field of view of
23$^{\circ} \times$ 13$^{\circ}$. The HDTV digital imagery has 1920
(horizontal) $\times$ 1035 (vertical) pixels that result in 6 times
higher resolution than the NTSC/PAL standard video conversion
system. The recording rate was 30 frames (60 fields) per
second. Spectroscopic observations were performed by the II-HDTV system
equipped with a reflection grating, which is 500 grooves mm$^{-1}$,
blazed at 330 nm, manufactured by the Richardson Grating Laboratory.

\begin{figure}[h]
\epsscale{1.00}
\plotone{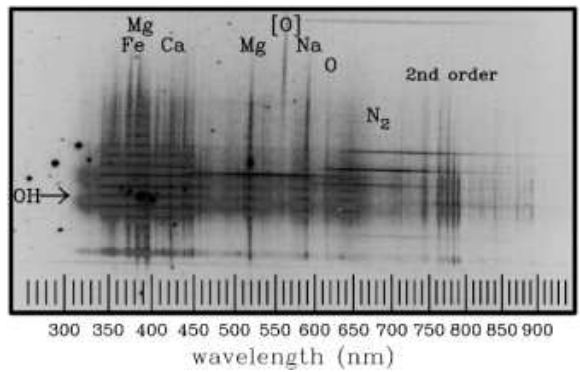}
\caption{Raw data of first and second order spectrum of 2001 Leonid
fireball at 18:58:20 UT on 2001 November 18. This image (with a field of
view of 23$^{\circ} \times$ 13$^{\circ}$) is composed of 15 consecutive
frames during a total duration of 0.5 s. The meteor moved from top to
bottom in this image. The dispersion direction is from left to right,
and parts of the second and third order spectra are on the right.}
\label{fig1}
\end{figure}

Background stars were removed by subtracting a median frame shortly
before the appearance of the meteor. After flat-fielding and averaging
the meteor spectrum, the wavelength was determined carefully by means of
numerous well-known atomic lines in the meteor emission. The effective
spectral sensitivity of the instrument, including atmospheric extinction
during the observations, was constructed by measuring spectra of bright
stars in the observing field. Its sensitivity covered the wavelength at
300-700 nm, with the maximum at $\sim$430 nm. The resulting dispersion
of the spectrum is 0.37 nm pixel$^{-1}$, and FWHM = 2.5 nm in the first
order. Since no order-sorting filter was used, it turns out that the
first-order spectrum was mixed with the second-order spectrum in the
wavelength longer than 600 nm. Details of the instrument and the first
results of the II-HDTV spectrum of 1999 Leonids were described in
\citet{abe00}.

Ozone in the stratosphere strongly absorbs below 290 nm, preventing the
UV light from reaching the Earth's surface. In order to prevent air
extinction owing to mainly aerosol scattering in the UV wavelength below
380 nm, the spectroscopic observation was performed at a high-altitude
site in the Nobeyama Radio Observatory of the National Astronomical
Observatory of Japan (N35.$^{\circ}$93, E138.$^{\circ}$48, altitude =
1340 m). Thanks to excellent observing conditions, clear weather, and
strong meteor storm activity during the Leonid maximum, its peak
activity was well observed around 18:16 UT on 2001 November 18 with the
zenithal hourly rate of 3730, based on reports of the Nippon Meteor
Society and the International Meteor Organization \citep{arlt01}.

\section{Results}

In Figure \ref{fig1}, we show a clear spectrum of a Leonid meteor
fireball, which was obtained at 18:58:20 UT on 2001 November 18, within
a dust trail ejected by comet 55P/Tempel-Tuttle in 1866
\citep{mcnaught99}. Assuming the Leonid radiant ($\alpha$ =+153
$^{\circ}$, $\delta$ = +22$^{\circ}$) and velocity (72 km s$^{-1}$), the
meteor distance $R$ and its altitude $H$ at each frame were
inferable. The altitudes for the fireball entering our field of view and
for it disappearing were $H$ = 108.1 km ($R$ = 136.8 km) and $H$ = 80.1
km ($R$ = 124.6 km), respectively.

From a comparison between meteor emission lines and the field-star
spectrum, a maximum brightness of -- 4 visual magnitude at the standard
range of 100 km was derived, which corresponded to a photometric
meteoroid mass of $\sim$1.8 g and a diameter of $\sim$15 mm by assuming
a density of 1.0 g cm$^{-3}$ for the Leonid meteoroids. All spectral
luminosities were normalized at 100 km altitude above the observer. In
the next section, we shall focus on the best spectrum at $t$ = 0.434 s
and $H$ = 84.1 km among these sequences.

First, in order to consider atomic lines only, we focus on emission
lines in the UV-VIS region below 450 nm, where the measurement of the
sensitivity calibration is expected to be the brightest. Assuming the
local thermal equilibrium (LTE), the atomic synthetic spectrum was
computed by adjusting five parameters: temperature $T$ and the column
densities of four atoms (Fe, Mg, Ca, and Na). We made sure that all
other possible atoms, such as H, N, O, Al, Si, Ti, Cr, Mn, Co, of Ni,
could be negligible in this wavelength region because these minor
emissions blended into strong emission lines owing to low-resolution
spectroscopy and because no significant contribution was
identified. This analytical method was described in
\citet{borovic93}. In general, meteor spectra consist of two components
at different temperatures \citep{borovic94}. A typical temperature of
the ``main (warm) component,'' which contains most of the above spectral
lines, is $T~\sim$ 4500 K. The ``second (hot) component'' is excited at
$T~\sim$ 10,000 K and consists of a few ionized elements such as
\ion{Ca}{2} and \ion{Mg}{2}.

\begin{figure}[h]
\epsscale{1.00}
\plotone{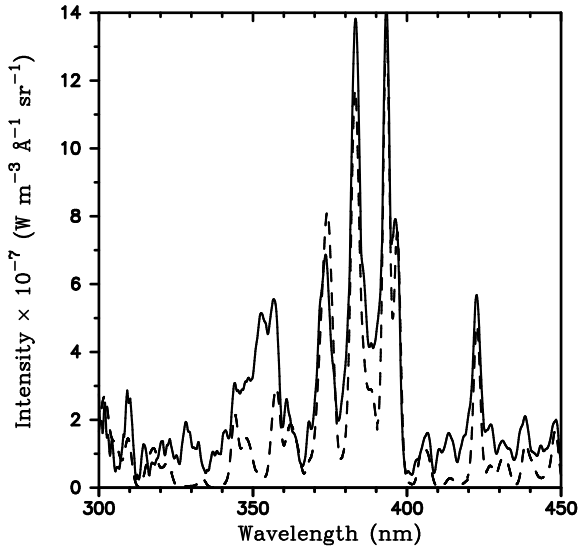}
\caption{Observed spectrum ($solid line$) compared with synthetic
spectrum considering only atoms ($dashed line$). The synthetic spectrum
consists of \ion{Mg}{1} at 383.8 nm and \ion{Ca}{2} at 393.4 and 396.8
nm, \ion{Ca}{1} at 422.7 nm and \ion{Mg}{2} at 448.1 nm with numerous
iron lines. There are unexplained emissions at $\sim$350, $\sim$330 and
$\sim$310 nm.}
\label{fig2}
\end{figure}

Although overlaps of numerous iron lines prevented us from determining
the precise temperature owing to rather low-resolution spectroscopy, LTE
temperatures of 4100-4700 K for the main spectrum of this Leonid meteor
have resulted in all spectrum frames, except saturated frames. Thus, we
applied typical temperatures of 4500 and 10,000 K for its warm and hot
component spectra, respectively. Figure \ref{fig2} shows the comparison
between observed and synthetic atomic spectra. The resulting column
density of \ion{Fe}{1} atoms was 2 $\times$ 10$^{15}$ cm$^{-2}$, and we
derived the following atomic ratios in the radiation gas:
\ion{Mg}{1}/\ion{Fe}{1} = 11, \ion{Ca}{1}/\ion{Fe}{1} = 0.1, and
\ion{Na}{1}/\ion{Fe}{1} = 0.03. Higher than chondritic Mg/Fe ratio was
also detected in a 2001 Leonid fireball \citep{borovic04}.

\begin{figure}[h]
\epsscale{1.00}
\plotone{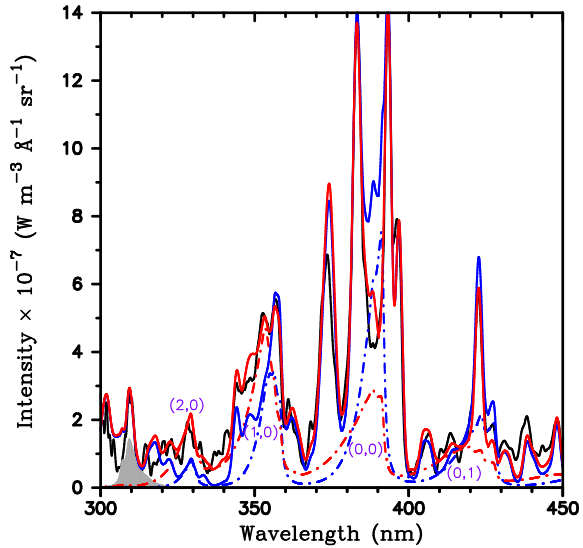}
\caption{Observed spectrum ($black line$) compared with synthetic
spectrum considering atoms and molecules of N$_{2}^{+}$(1--) with a
temperature of 10,000 K ($red line$) and 4400 K ($blue line$). The
dash-dotted lines indicate N$_{2}^{+}$(1--) at the appropriate
temperature. The gray area near 310 nm shows OH $A$--$X$ bands.}
\label{fig3}
\end{figure}

After the comparison between observed and synthetic spectra in Figure
\ref{fig2}, we found that these atomic lines could not help us identify
some unknown bands around 350 and 330 nm. These unknown bands clearly
appeared from $t$ = 0.167 s ($H$ = 100.1 km) and suddenly disappeared
after $t$ = 0.467 s ($H$ = 82.1 km). The 350 nm excess was particularly
strong. In order to explain these enhancements in accuracy, the SPRADIAN
numerical code considering with Franck-Condon factors, which can produce
a molecular spectrum at the appropriate temperature and density, was
used \citep{fujita97}. As a result, the best account for two excess
bands at 330 and 350 nm was found to be the ``first negative $B$--$X$''
band of the molecular nitrogen ion N$_{2}^{+}$(1--). Figure \ref{fig3}
shows the model spectrum of N$_{2}^{+}$(1--) with four bands heads
(329.3, 353.4, 391.4, and 427.8 nm) caused by different vibrational
states of (v',v'')=(2,0), (1,0), (0,0), and (0,1), respectively. The
model spectrum is wonderfully in complete agreement with the
observational spectrum in the UV range from 320 to 360 nm. This finding
is the first detection of the N$_{2}^{+}$ $B$--$X$ molecule in the UV
meteor spectrum. The final values of number densities and chemical
abundances for this meteor spectrum are summarized in Table~\ref{tbl-1}
and ~\ref{tbl-2}.

\section{Discussion}

The presence of the molecular nitrogen ion N$_{2}^{+}$(1--) in meteor
spectra was first suspected by
\citet{millman71}. \citet{mukhamednazarov77} reported the detection of
the (0, 1) and the (0, 0) bands in the spectra of faint 3-5 mag meteors
with the aid of intensified TV cameras. However, since the resolution
was low ($\sim$5 nm), the presence of these bands could not be separated
from the strong emissions of \ion{Ca}{1} at 422.7 and \ion{Ca}{2} at
393.4 nm. Also, the (0, 0) band was almost at the edge of the their
instrumental sensitivity. Figure \ref{fig2} shows a clear contribution
of \ion{Ca}{1} at 422.7 nm and many iron lines that overlap with the (0,
1) band. In addition to that, these (0, 0) and (0, 1) bands of
N$_{2}^{+}$(1--) were never found in fireballs. That is, previous
reports could not identify the N$_2^{+}$(1--) because it was difficult
to detect these bands clearly in the VIS region because of the overlaps
with strong Ca and Fe emissions.

On the other hand, \citet{jenn04a} found excessive emissions between 770
and 840 nm with the maximums centered at $\sim$789 and $\sim$815 nm,
which could be caused by the ``first negative $A$--$X$'' band of the
molecular nitrogen ion N$_{2}^{+}$ Meinel bands. The evidence of a
N$_{2}^{+}$ Meinel system was identified in two fireballs (with
magnitudes of -- 1 and -- 7) obtained during the Leonid meteor shower in
2001 and 2002 Leonid Multi-Instrument Aircraft Campaign (MAC;
\citep{jenn02a}. Because their unintensified slitless CCD spectrograph
could provide a high spectral resolution with the precise determination
of wavelength in the near-IR region, these must be reliable findings. In
their research, the LTE abundances ($T$ = 4400 K) of
[N$_{2}^{+}$]/[N$_{2}$] for the -- 1 and -- 7 mag fireballs were
estimated at 5 $\times$ 10$^{-7}$ and 2 $\times$ $10^{-9}$,
respectively. In addition to this, a tentative identification of the
fist negative N$_{2}^{+}$ $B$--$X$ (0, 0) band in the VIS region was
proposed \citep{jenn04a,jenn04b}.

\begin{figure}[h]
\epsscale{1.00}
\plotone{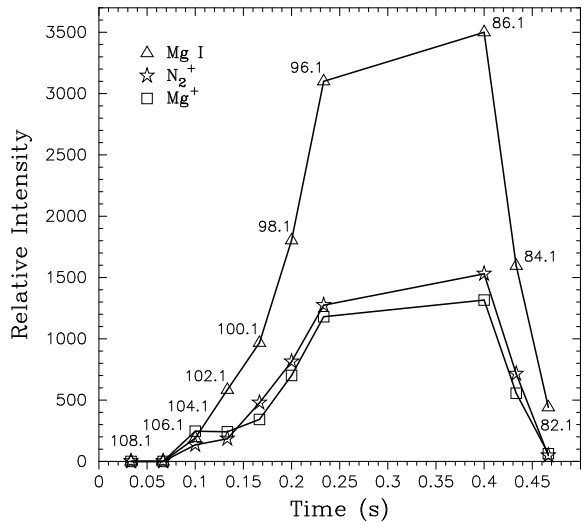} 
\caption{Light curves for N$_{2}^{+}$ (353 nm) \ion{Mg}{2} (448 nm), and
\ion{Mg}{1} (518 nm). Time in seconds is on the horizontal axis. The
relative intensity in an arbitrary linear scale is on the vertical axis.
Altitude (in units of kilometers) at an appropriate time is
indicated. Saturated spectra during the meteor flare ($t$ = 0.276 -
0.400 s, $H$ = 94.1 - 86.1 km) are omitted.}
\label{fig4}
\end{figure}

Although the spectral resolution in our observation was about an order
less than Jenniskens' results, we could take full advantage of the
sensitivity in the UV region below 380 nm. It is obvious that the
spectrum enhancement around 350 and 330 nm can be explained by band
heads of N$_{2}^{+}$ $B$--$X$ (1, 0) at 353.4 nm and (2, 0) at 329.3
nm. The second positive bands of the neutral N$_{2}$ molecule were not
identified in the UV-VIS range, which should contribute as a
background. We inferred an upper limit of neutral N$_{2}$ of $\sim$1.0
$\times$ 10$^{13}$ cm$^{-3}$ by assuming an LTE temperature of 4500 K
\citep{jenn00,jenn04a}. The best-fit calculation mixed with atomic lines
leads to an N$_{2}^{+}$(1--) vibrational temperature of 10,000 K. If we
assume an N$_{2}^{+}$ temperature of 4500 K, the estimated abundance of
[N$_{2}^{+}$]/[N$_{2}$] leads to $\sim$ 1 $\times$ 10$^{-7}$, which is
consistent with the results from \citet{jenn04a}. However, the
best-fitted spectrum clearly proves that N$_{2}^{+}$(1--) belongs to the
``hot component'' of $T$ = 10,000 K. Figure \ref{fig3} indicates the
synthetic spectrum assuming $T$ = 4400 K, which is clearly different
from the observed spectrum. Furthermore, another piece of possible
evidence for this hot component being assigned to N$_{2}^{+}$ is that
the light curves of N$_{2}^{+}$ emission is more similar to ionized Mg
at 448.1 nm which belongs to hot component, rather than neutral Mg at
517.8 nm which belongs to warm component (Fig. \ref{fig4}).

The N$_{2}^{+}$(1--) system was surprisingly strong; its total flux
between 300 and 450 nm was 1.44 $\times$ 10$^{-4}$ (W m$^{-2}$
sr$^{-1}$) even when the calculated number density of N$_{2}^{+}$
molecules was extremely small as 1.55 $\times$ 10$^{5}$ cm$^{-3}$; i.e.,
[N$_{2}^{+}$]/[N$_{2}$] = 1.5 $\times$ 10$^{-8}$. Few reports have
clarified the UV (300-350 nm) meteor spectra in the past
\citep{harvey73a,harvey73b,carbary02,jenn02b}, and strong features
related to N$_{2}^{+}$(1--) had never been reported before. On the other
hand, similar features around 350 nm can be seen in spectra of 1999
Leonid meteoroids obtained by \citet{rairden00} and in our fireball
spectra observed in 2002 Leonids; both were observed during the Leonid
MAC. N$_{2}^{+}$(1--) also could be observed in a shock layer of arc
plasma heated by the reflected shock and by reentry spacecraft such as
the Space Shuttle orbiter \citep{viereck92}. The flow in the head and
wake regions of a hypersonic object, such as a reentry capsule, tends to
be in a thermochemical nonequilibrium state. The most likely scenario of
the induced N$_{2}^{+}$(1--) in the meteoroid will result in the effect
of large dimensions of high-temperature regions just ahead and behind
the meteoroid caused by the large meteoroids' vapor cloud. The reentry
speed from interplanetary space is more than 12 km s$^{-1}$ at the
altitude of 100 km, which is enough velocity for producing
N$_{2}^{+}$(1--) suggested from laboratory experiments
\citep{keck59}. Therefore, the reentry capsules (meter-size meteoroids)
of $Genesis$ (solar wind sample return), $Stardust$ (cometary dust
sample return), and $Hayabusa$ (asteroidal material sample return)
directly from interplanetary space, which will return to the Earth in
2004 September, 2006 January, and 2007 June, respectively, will provide
us with good opportunities to conduct artificial fireball spectroscopy
tests in the future \citep{yano04}.

\begin{table}
\begin{center}
\caption{Chemical composition of the Leonid meteor spectrum 
($t$ = 0.434 s, $H$ = 84.1 km) at 18:58:20 UT on 2001 November 18}
\label{tbl-1}
\begin{tabular}{ccc}
\tableline\tableline
             & Number Density & Total Flux  \\
Emission     &  (cm$^{-3}$)  & (10 $^{-5}$ W m$^{-2}$ sr$^{-1}$) \\ \tableline
N$_2^+$      &    1.55 $\times$ 10$^{5}$  ($T$=10,000 K)& 14.4   (300-450 nm) \\
N$_2$        & $<$ 1.0 $\times$ 10$^{13}$ ($T$=4500 K) & 1.29   (300-450 nm) \\
OH           & $>$ 7.9 $\times$ 10$^{7}$  ($T$=4500 K) & 1.59   (300-330 nm) \\
\ion{Fe}{1}  & 5.5 $\times$ 10$^{13}$    &  \\
electron     & 7.6 $\times$   10$^{13}$  &  \\
\tableline
\end{tabular}
\end{center}
\end{table}

\begin{table}
\begin{center}
\caption{Chemical abundances}
\label{tbl-2}
\begin{tabular}{ccc}
\tableline\tableline
        &           &  T  \\
Element & Abundance & (K) \\ \tableline
Mg/Fe & 11  & 4500\\
Ca/Fe & 0.1 & 4500\\
Na/Fe & 0.03& 4500\\
N$_2^+$/N$_2$ & $>$ 1.5 $\times$ 10$^{-8}$ & \\
N$_2^+$       &                            & 10,000 \\
N$_2$         &                            & 4500 \\
N$_2^+$/N$_2$ & $\sim$ 1 $\times$ 10$^{-7}$& \\
N$_2^+$       &                            & 4500 \\
N$_2$         &                            & 4500 \\
\tableline
\end{tabular}
\end{center}
\end{table}

\acknowledgments

The observations were carried out by M. Sugimoto (NMS), N. Fujino (Tokyo
University), T. Kasuga (NINS/NAOJ), K. Oka, M. Watanabe and A. Yamada
(Japan Women's University). The authors would like to thank M. Inoue
(Nobeyama Radio Observatory, NAOJ/NINS) and H. Ando (Subaru Observatory,
Hawaii, NAOJ/NINS) for their observational support. The authors are also
deeply indebted to K. Fujita (ISTA/JAXA) and Y. Hirahara (Nagoya
University) for their advice in the molecular calculation. This research
was supported by the National Astronomical Observatory of Japan of the
National Institutes of Natural Sciences, the Institute of Space and
Astronautical Science of the Japan Aerospace Exploration Agency, the
Institute of Physical and Chemical Research, and the National Institute
of Information and Communications Technology. This study is carried out
as a part of ``Ground-based Research Announcement for Space
Utilization'' promoted by the Japan Space Forum. S. A. is supported by
JSPS Postdoctoral Fellowships for Research Abroad.

\appendix

\end{document}